\documentclass[namedreferences]{SolarPhysics}
\usepackage[optionalrh]{spr-sola-addons} 
\usepackage{graphicx}        
\usepackage{color}           
\usepackage{url}             

\newcommand{\etal}{{\it et al.}}

\newcommand{\apj}      {{\it Astrophys. J.}}

\newcommand{\solphys}  {{\it Solar Phys.}}

\begin{document}

\begin{article}

\begin{opening}
\title{Super Fast and Quality Azimuth Disambiguation \\ {\it Solar Physics}}

\author{G. V.~\surname{Rudenko}$^{1}$\sep
               S. A.~\surname{Anfinogentov}$^{1}$\sep
       } \runningauthor{G.V.
Rudenko and S.A. Anfinogentov} \runningtitle{Super Fast and
Quality Azimuth Disambiguation}

   \institute{$^{1}$ Institute of Solar-Terrestrial Physics SB RAS, Lermontov St. 126,
Irkutsk 664033, Russia
                     email: \url{rud@iszf.irk.ru} email: \url{anfinogentov@iszf.irk.ru}\\}
\begin{abstract}
The paper presents the possibility of fast and quality azimuth
disambiguation of vector magnetogram data regardless of location
on the solar disc. The new Super Fast and Quality (SFQ) code of
disambiguation is tried out on well-known models of
\inlinecite{Metcalf2}, \inlinecite{Leka1} and an artificial model
based on observed magnetic field in AR 10930 \cite{Rudenko}. We
make comparison of Hinode SOT SP vector magnetograms of AR 10930
disambiguated with three codes: SFQ, NPFC \cite{Georgoulis}, and
SME \cite{Rudenko}. We illustrate the SFQ disambiguation on
SDO/HMI full disk magnetic field observations.  The preliminary
examination indicates that the SFQ algorithm provides better
quality than NPFC and is comparable to SME. In contrast to other
codes, SFQ supports relatively high quality which does not depend
on the distance to the limb (unlike all other algorithms the
presented one remain efficient even  when the magnetigram position
is very close to the limb).
\end{abstract}
\end{opening}
\section{Introduction}
Azimuth disambiguation of the transverse field in vector
magnetograph data is a key problem determining reliability of the
physical researches utilizing the knowledge of full vector of the
photospheric magnetic field. Apparently complete and exact
solution to this problem is hardly possible because of high level
of noise in transverse field data and impossibility of spatial
resolution of real pattern of the magnetic field (whose thin
structure is significantly less than the spatial resolution of
modern magnetographs). In this connection, the main requirement
for algorithms of the azimuth disambiguation is to reach
invariably high quality that could describe adequately (with
minimum distortions) the physically significant real structural
elements of magnetic configurations. Also of importance is the
code efficiency with regions regardless of their location on the
solar disc; this allows us to use data on all available
measurements, including those near the limb. To perform on-line
processing of continuously incoming data, the final preparation
time should be comparable to time of measurement data acquisition.
Currently, the best codes providing invariably high quality and
meeting the requirements for physical investigation are the codes
based on the simulated annealing algorithm (different variants of
the "minimum energy" method (ME \inlinecite{Metcalf1};
\inlinecite{Leka2}, \inlinecite{Rudenko}). Quality of the code
outcome is significantly higher than the results of other the
results available numerous codes \cite{Metcalf1}. Unfortunately,
ME codes are among the most long-term: the required time goes up
with increasing number of nodes of data grid. Therefore they are
not suitable for real-time processing of continuous data flow with
high spatial resolution. The SDO/HMI instrument, for instance,
generates vector magnetogram with resolution of 4096x4096 once
every 12 minutes.

A fast NPFC algorithm \cite{Georgoulis} is now examined in order
to be used in data flow processing, since it represents the best
compromise between quality and time required for processing.
Quality of the NPFC algorithm is ranked second among well-known
methods, though it is inferior to that of ME codes. It does not
always show invariably satisfactory results in different real
configurations of magnetic regions, particularly in those near the
limb.

In this paper, we present examples of the azimuth disambiguation
of model and real ambiguous magnetograms with the use of three
codes: the new Super Fast and Quality (SFQ), NPFC, and SME
\cite{Rudenko}.  We show that quality of the SFQ outcome far
surpasses that of NPFC. Besides, SFQ is much faster than NPFC.
Moreover, time for the azimuth disambiguation of
vector-magnetograph data with any spatial resolution (including
SDO/HMI measurements of the full disc) may be significantly
reduced due to the SFQ parallelization into several processes.
This suggests potential on-line processing of current data flow.
\section{The method}
The SFQ method is a two-step processing system. Step 1 involves
preliminary azimuth disambiguation, using special metric (grid
difference metric) relying on reference information of the
potential field. Step 2 (cleaning) comprises application of
smoothing masks in several scales. Our solution does not use
random search (like in ME) or convergent iterations (like in NPFC)
that's why it is very fast.
\subsection{Step 1}
The key point of the method is application of the "grid difference
metric" defining measure of difference between the initial
ambiguous field ${\bf B}^{amb}$ and the potential (reference)
field ${\bf B}^{ref}$. Metric is constructed for each node $I$,
$j$ of the magnetogram grid as follows:

\begin{eqnarray}\label{eq:1}
g^{ij}\left( {\bf B}_{\perp }^{amb}\right)
=\sqrt{\sum_{s=0}^{1}\sum_{t=0}^{1}\left[ \Delta _{s}\left(
B_{t}^{amb}-B_{t}^{ref}\right) \right] ^{2}};\nonumber\\
\Delta _{0}=f=f^{i+1,j}-f^{i,j},\Delta _{1}=f=f^{i,j}-f^{i,j+1};\\
B_{0}=B_{x},B_{1}=B_{y}.\nonumber
\end{eqnarray}
Metric (\ref{eq:1}) is then used as a conditional mask to
determine common sign for transverse components of the preliminary
field $B^{Step\_1}$:
\begin{equation}\label{eq:2}
\left( {\bf B}_{\perp }^{Step\_1} \right)^{i,j}=\left\{%
\begin{array}{c}
  \left( {\bf B}_{\perp }^{amb} \right)^{i,j} \\
  -\left( {\bf B}_{\perp }^{amb}\right)^{i,j} \\
\end{array}%
 \right|
 \left.%
\begin{array}{c}
  if \quad g^{i,j}\left( {\bf B}_{\perp }^{amb}\right) \leq g^{i,j}\left( -{\bf B}_{\perp }^{amb}\right)\\
  if \quad g^{i,j}\left( {\bf B}_{\perp }^{amb}\right) > g^{i,j}\left( -{\bf B}_{\perp }^{amb}\right) \\
\end{array}%
 \right\}
\end{equation}
According to definition of metric (\ref{eq:1}), its application is
valid only for nodes with locally continuous ${\bf B}^{amb}$
(i.e., random distribution over azimuth direction nodes fails for
${\bf B}^{amb}_{\perp }$). Consequently, requirement that initial
distribution of the transverse field be locally continuous in most
nodes is a necessary condition. This requirement is easily
satisfied, for instance, if a component of the transverse field
($x$ or $y$) is assumed to be everywhere positive.  The potential
reference field can be obtained using the standard method for the
FFT extrapolation over the longitudinal component $B_z$. In our
case, we used the potential extrapolation in quasi-spherical
geometry \cite{Rudenko} and subsequent extrapolation of the field
into nodes of the initial magnetogram grid.

We think that the principle of construction of metric $g$
(\ref{eq:1}) is similar to the metric used in the most effective
ME method:
\begin{equation}\label{eq:3}
E=\left| j_z \right|+\left| \nabla_\perp \cdot {\bf B} +
\partial_z B^{pot}_z \right|\equiv \left| j_z -j^{pot}_z \right|+\left| \nabla_\perp \cdot {\bf B} -\nabla_\perp \cdot {\bf B}^{pot}\right|
\end{equation}
Indeed, close inspection of the equation shows that, formally,
terms in the right part of identity (\ref{eq:3}) are the
particular linear combinations of differences (\ref{eq:1}). For
simplicity of comparison, we will confine ourselves to the case of
flat approximation and $B_l=B_z$. We attach certain physical
meaning to summands of (3). The first summand in (\ref{eq:3})
provides minimisation of vertical current. This term is
responsible for nodes in which local continuity of the transverse
ambiguous field is disturbed. In other nodes, it does not depend
on azimuth direction of the transverse field. The second summand
in (\ref{eq:3}) is the divergence simulation; it reflects
correlation of some differences between the test and reference
fields (i.e., it performs function analogous to (\ref{eq:1}). It
is thus reasonable to suppose that the main positive function of
the azimuth selection in both cases is performed by differential
relations between the test and reference fields. We do not assign
value of a geometric object's element (similar to the value of
field derivatives making up tensor components of vector field
derivatives) to each difference $\Delta _{s}f$ in (\ref{eq:1}). We
consider these differences as formal functions of two near points,
reflecting local coupling between the test and reference fields.
This approach allows us to consider all magnetogram nodes as
equivalent (regardless of their location in the visible part of
the photosphere) and relieves us of having to use additional grids
and geometric transformations. Notice that the NPFC method (which
is referred to as nonpotential) uses approximation $\frac{\partial
B_{z}}{\partial z}=\frac{\partial B_{z}^{pot}}{\partial z}$
\cite{Metcalf2} when deducing equations for current component
$B_c$ in indirect form. This condition implies differential
reference relation with the potential field.
\subsection{Step 2}
The necessity to clean the preliminary magnetogram after Step 1 is
caused by the peculiar features of our application of metric
(\ref{eq:1}). Let us suppose that we apply Step 1 to a noise-free
magnetogram of the potential field. In this case, we would
probably observe the following: solution to the disambiguation
problem is rigorous in the main continuum of points in whose
vicinities initial distribution of the transverse field is
continuous. Random chains of isolated contrast (bad) pixels on
transverse component images would be observed only on lines of the
local discontinuity of data. Almost all these bad pixels may be
cleaned at one go through comparing with the smoothed transverse
field $\overline{{\bf B}}_\perp$:
\begin{equation}\label{eq:4}
\left( {\bf B}_{\perp }^{n+1} \right)^{i,j}=\left\{\left. %
\begin{array}{c}
  \left( {\bf B}_{\perp }^{n} \right)^{i,j} \\
  -\left( {\bf B}_{\perp }^{n}\right)^{i,j} \\
\end{array}%
 \right.\left|
\begin{array}{c}
  if \quad \left( {\bf B}_{\perp }^{n}\cdot\overline{{\bf B}}_\perp^n\right)^{i,j} \leq 0\\
  if \quad \left( {\bf B}_{\perp }^{n}\cdot\overline{{\bf B}}_\perp^n\right)^{i,j} > 0 \\
\end{array}%
 \right.\right\}
\end{equation}

\begin{figure}
  \includegraphics[width=2.8\linewidth]{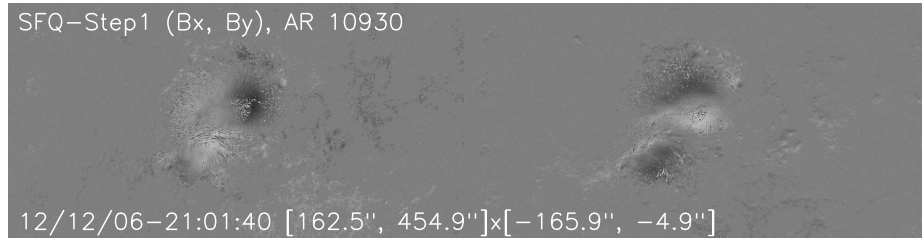}\\
  \caption{Transverse components of the magnetic region AR 10930 near the disc centre after Step 1 processing.}\label{fig:1}
\end{figure}

\begin{figure}
  \includegraphics[width=1.4\linewidth]{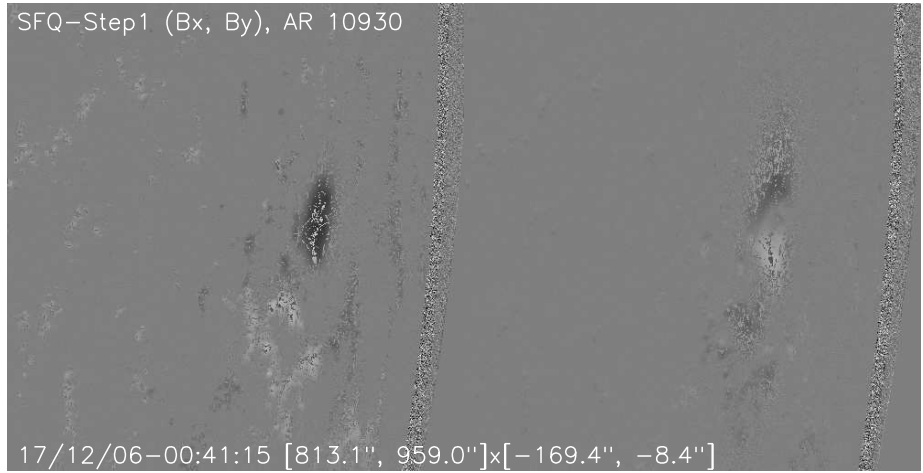}\\
  \caption{Transverse components of the magnetic region AR 10930 near the limb after Step 1 processing.}\label{fig:2}
\end{figure}

Typical distribution of bad pixels after Step 1 processing of real
magnetograms is well exemplified by images of transverse
components $(B_x, B_y)$ of Hinode/SOT SP data (Level 2) in the
magnetic region AR10930 for two its locations - near the disc
centre (Fig. \ref{fig:1}) and near the limb (Fig. \ref{fig:2}).

Figures show that the general pattern of the transverse field is
satisfactorily neat already after Step 1 processing regardless of
bad pixels; noteworthy is the fact that the outcome quality near
the limb is comparable to that near the centre. This suggests that
the chosen metric (\ref{eq:1}) is quite effective, irrespective of
proximity to the limb. Close inspection of images in Figures
\ref{fig:1} and \ref{fig:2} shows that there are many bad
microfragments (groups of adjacent bad pixels). It is obvious that
single application of (\ref{eq:4}) for many of such microfragments
may be insufficient. On the other hand, it is reasonable to expect
that multiple application of (\ref{eq:4}), given relevant
smoothing parameters, may lead to the "collapse" of fragments due
to the motion of their boundaries towards the decrease in absolute
magnitude of the transverse field. In practice, most types of bad
fragments have the necessary feature: closed fragments, as a rule,
collapse (towards the necessary side), whereas open fragments
(part of the boundary in the weak noisy field is not seen)
transfer their boundaries to the noise region.

We have conducted numerous tests and selected the following type
of cleaning (Step 2). Two types of smoothing procedures are used
for cleaning. In terms of the IDL code, they have the form: \\
a)  Smoothig(s) - sBx=smooth(Bx,s,$\backslash$edge)-Bx/float(s\symbol{"5E}2) \& sBy=smooth(By,s,$\backslash$edge)-By/float(s\symbol{"5E}2) \\
b)  Median(s) - sBx=median(Bx,s,$\backslash$even) \& sBy=median(Bx,s,$\backslash$even) \\
Here, $s\times s$ is the size of smoothing window. In this form,
smoothing in each node of the grid is performed only in adjacent
nodes, exclusive of node values. This modification enhances
features of the collapse. Either of smoothing types a) or b) for
the chosen parameter $s$ is repeatedly used with subsequent
procedure (\ref{eq:4}) in the cycle with the exit condition:
number of iterations reaches 300 or number of modified pixels is
less than 5 or 0.01 \% of the total amount of pixels.

The final cleaning (Step 2) is the following set of consecutive
cycles: \\
Loop1 - \emph{Median}(3); \\
Loop2 - \emph{Smoothig}(19); \\
Loop3 - \emph{Smoothig}(9); \\
Loop4 - \emph{Smoothig}(5); \\
Loop5 - \emph{Smoothig}(3).\\
We have used this scheme for all the examples of calculations of
model and real magnetograms given before \footnote{Many
configurations of "cleaning" mode are possible. Probably some of
them can provide better results. So the "cleaning" version
described in this paper is most likely not final and may be
improved in the future. The latest version of the SFQ code
implemented in IDL language is available at
\url{http://bdm.iszf.irk.ru/sfq_idl}}.

\section{Model tests}
\subsection{Known models} \label{sect:known_models}

To compare SFQ with other codes, we have processed all ambiguous
magnetograms\footnote{The magnetograms are available on our web
page \url{http://bdm.iszf.irk.ru/SFQ_Disambig}} of known models
used by \inlinecite{Metcalf2} and \inlinecite{Leka1} to estimate
the majority of known methods for the azimuth disambiguation:
\begin{itemize}
    \item \url{http://www.cora.nwra.com/AMBIGUITY_WORKSHOP/2005/DATA_FILES/Barnes_TPD7.sav}
    \item \url{http://www.cora.nwra.com/AMBIGUITY_WORKSHOP/2005/DATA_FILES/fan_simu_ts56.sav}
    \item \url{http://www.cora.nwra.com/AMBIGUITY_WORKSHOP/2006_workshop/TPD10/TPD10a.sav}
    \item \url{http://www.cora.nwra.com/AMBIGUITY_WORKSHOP/2006_workshop/TPD10/TPD10b.sav}
    \item \url{http://www.cora.nwra.com/AMBIGUITY_WORKSHOP/2006_workshop/TPD10/TPD10c.sav}
    \item \url{http://www.cora.nwra.com/AMBIGUITY_WORKSHOP/2006_workshop/FLOWERS/flowers13a.sav}
    \item \url{http://www.cora.nwra.com/AMBIGUITY_WORKSHOP/2006_workshop/FLOWERS/flowers13b.sav}
    \item \url{http://www.cora.nwra.com/AMBIGUITY_WORKSHOP/2006_workshop/FLOWERS/flowersc.sav}
\end{itemize}

 All processing results are presented on
\url{http://bdm.iszf.irk.ru/SFQ_Disambig/models.zip} by the set of
graphs of transverse components of the field (to make a visual
estimate) and by the corresponding set of digital IDL sav files (
to make quantitative assessment of vector magnetograms' quality).

\subsection{Answer vector model of AR 10930}\label{sect:answer}
The following type of testing of $\pi$-disambiguation methods
relies on the answer  vector model of the photospheric field of a
real magnetic region. Such a model can be easily obtained through
fixing vector components of the real component with removed
$\pi$-ambiguity (reference magnetogram) by the reference method
providing most reliable results. The answer model is the field
components of the reference magnetogram transformed into the
Carrington spherical coordinate system, with interpolation into
nodes of the uniform spherical grid. The resulting field is then
used to generate answer magnetograms simulating transit of the
selected magnetic configuration across the Sun's disc.
Transforming answer magnetograms into ambiguous ones and comparing
their processing effects, we can obtain quantitative assessment
for the method under study. Such an approach to modelling allows
us to present peculiarities of vector data and features of the
real magnetic structure to a great extent in the model. Degree of
quality of the quantitative assessment depends on quality of the
reference magnetogram. Disambiguation errors in the reference
magnetogram affect quality of the following tests. Tests in
section \ref{sect:known_models}  do not have this disadvantage; on
the other hand, there is no assurance that they properly represent
real peculiarities of data.

When making the answer vector model in our study, we used vector
data on the SOT/SP level 2 (12-17 December 2006) in AR 10930 and
the reference SME method of $\pi$ disambiguation \cite{Rudenko}.
Using the model obtained as basis, we simulated transit of a fixed
magnetic structure across the disc by a set of answer magnetograms
corresponding to moments of AR 10930 real measurement data. To
make quantitative assessment, we used standard parameters from
\inlinecite{Metcalf2} and \inlinecite{Leka1}:
\begin{eqnarray}\label{eq:metrics}
M_{area}=\#pixels(\Delta\theta =0)/\#pixels, \quad \Delta\theta
=0^0\lor \pm 180^0;\nonumber \\
M_{flux}=\sum\left(|B_n|_{\Delta\theta=0}\right)/\sum
|B_n|;\nonumber\\
M_{B_\perp >T}=\sum(B_\perp (s)_{\Delta\theta=0,B_\perp
>T})/\sum(B_\perp (s)_{B_\perp
>T})\\
M_{\Delta B}=\sum |{\bf B}(s)-{\bf B}(s)|/\#pixels ;\nonumber\\
M_{J_z}=M(a,s)_{J_z}=1-\frac{\sum(J_{n(answer)}-J_{n(solution)})}{2\sum
J_{n(answer)} } \nonumber.
\end{eqnarray}

Parameter values of $\pi$-disambiguation (\ref{eq:metrics}) of model magnetograms for three methods (SFQ, SME, and NPFC) are presented is Tables 1-6.

\begin{figure}
  \includegraphics[width=1.0\linewidth]{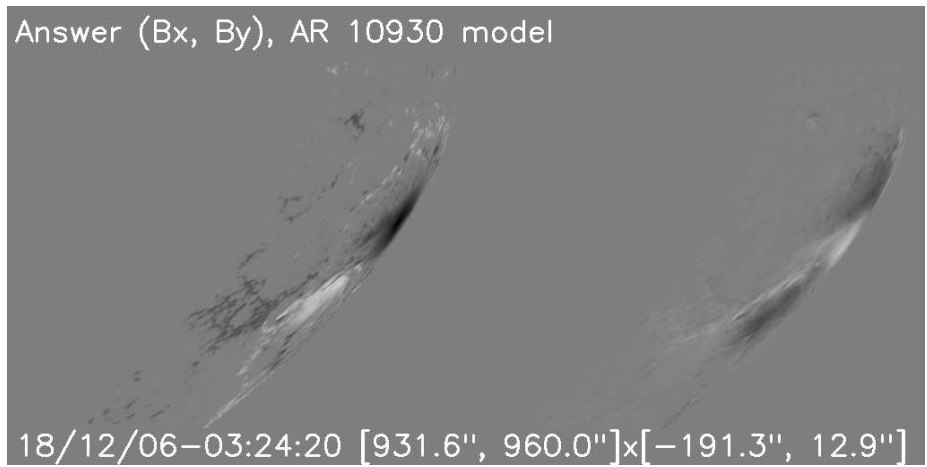}\\
  \caption{Limb answer magnetogram of the AR 10930 model. Lc=87.9}\label{fig:aranswer}
\end{figure}

\begin{figure}
  \includegraphics[width=1.0\linewidth]{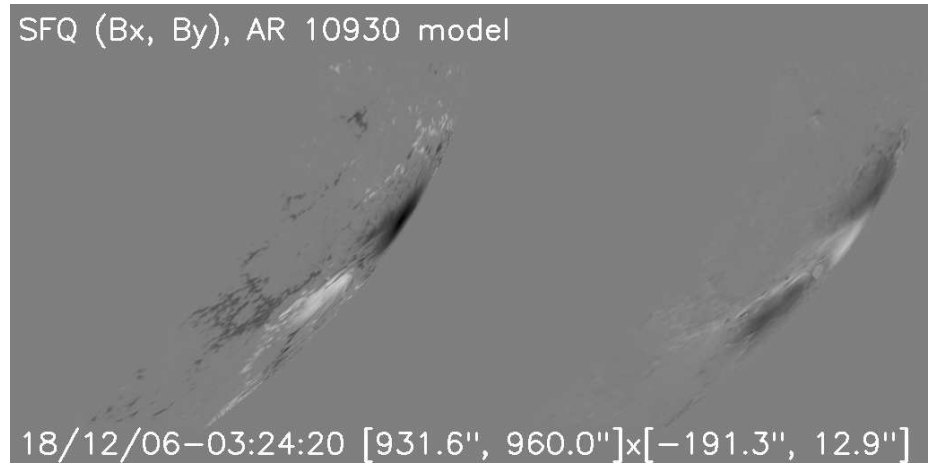}\\
  \caption{. Limb SFQ magnetogram of the AR 10930 model. Lc=87.9}\label{fig:aranswersfq}
\end{figure}

The first and second columns show position of magnetograms on the
disc: the first corresponds  to the time of real magnetograms with
the same position on the disc; the second demonstrates angular
distance from the disc centre to the centre of an arbitrarily
chosen region. Under close examination, we see that the SFQ
quality in these tables is closer to SME. NPFC demonstrates the
worst quality near the limb. When approaching the limb, most SFQ
parameters seem more preferable than SME. Besides, it is easily
seen that the SFQ quality all along the region transit is almost
always on the same high level, whereas results of other methods
deteriorate to a variable degree when approaching the limb. The
last line of the table does not correspond to the time of the real
magnetogram. Parameters of this line show the SFQ efficiency in
the artificial position of the region when major part of the
strong field stricture is behind the limb. Images of transverse
components of the answer magnetogram and SFQ magnetogram in this
case are presented in Fig. \ref{fig:aranswer} and
\ref{fig:aranswersfq}, respectively. This example proves that we
can apply the SFQ method for $\pi$ disambiguation of full-disc
magnetograms without significant distortions of magnetic
structures on the limb.

All other images corresponding to Tables 1-6 are available on:
\begin{description}
  \item[\url{http://bdm.iszf.irk.ru/SFQ_Disambig/Answer_AR10930_model.zip}] -- transverse components of answer magnetograms;
  \item[\url{http://bdm.iszf.irk.ru/SFQ_Disambig/SFQ_AR10930_model.zip}] -- transverse components of SFQ magnetograms;
  \item[\url{http://bdm.iszf.irk.ru/SFQ_Disambig/SME_AR10930_model.zip}] -- transverse components of SME magnetograms;
  \item[\url{http://bdm.iszf.irk.ru/SFQ_Disambig/NPFC_AR10930_model.zip}] -- transverse components of NPFC magnetograms
\end{description}

\section{Disambiguation of real magnetogram data}
\subsection{AR 10930}
Complete set of graphics files containing results of  $\pi$-disambiguation of real Hinode/SOT SP (Level~2) magnetogram data of AR 10930 by three methods is available on:
\begin{description}
  \item[\url{http://bdm.iszf.irk.ru/SFQ_Disambig/SFQ_AR10930.ZIP}] -- transverse components of SFQ magnetograms;
  \item[\url{http://bdm.iszf.irk.ru/SFQ_Disambig/SME_AR10930.ZIP}] -- transverse components of SME magnetograms;
  \item[\url{http://bdm.iszf.irk.ru/SFQ_Disambig/NPFC_AR10930.ZIP}] -- transverse components of NPFC magnetograms.
\end{description}

\begin{figure}
  \includegraphics[width=1.0\linewidth]{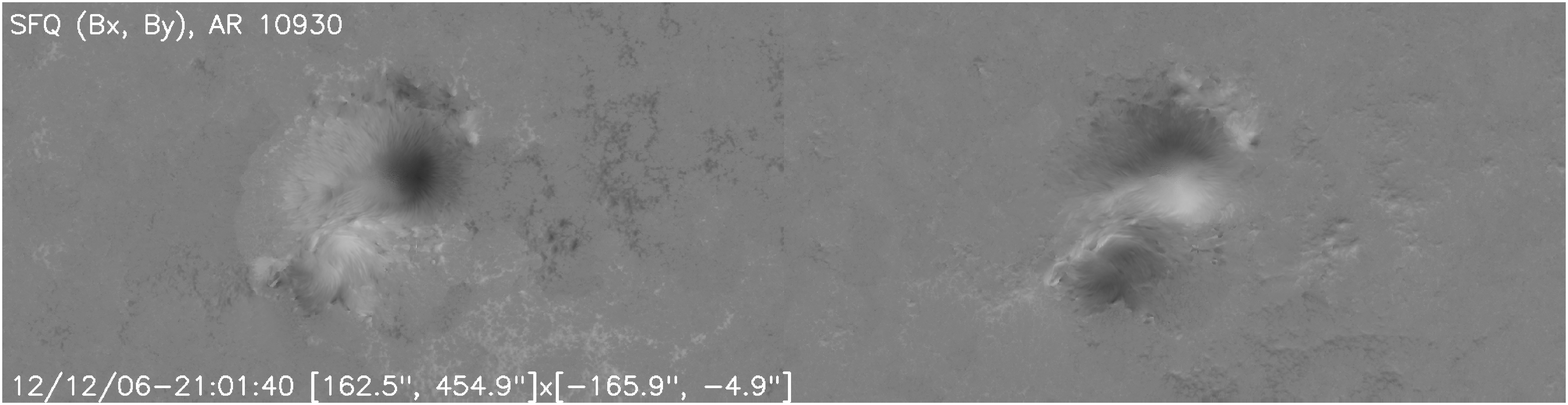}\\
  \caption{Near-Centre SFQ magnetogram of AR 10930. Lc=19.2}\label{fig:ar10930_1}
\end{figure}

\begin{figure}
  \includegraphics[width=1.0\linewidth]{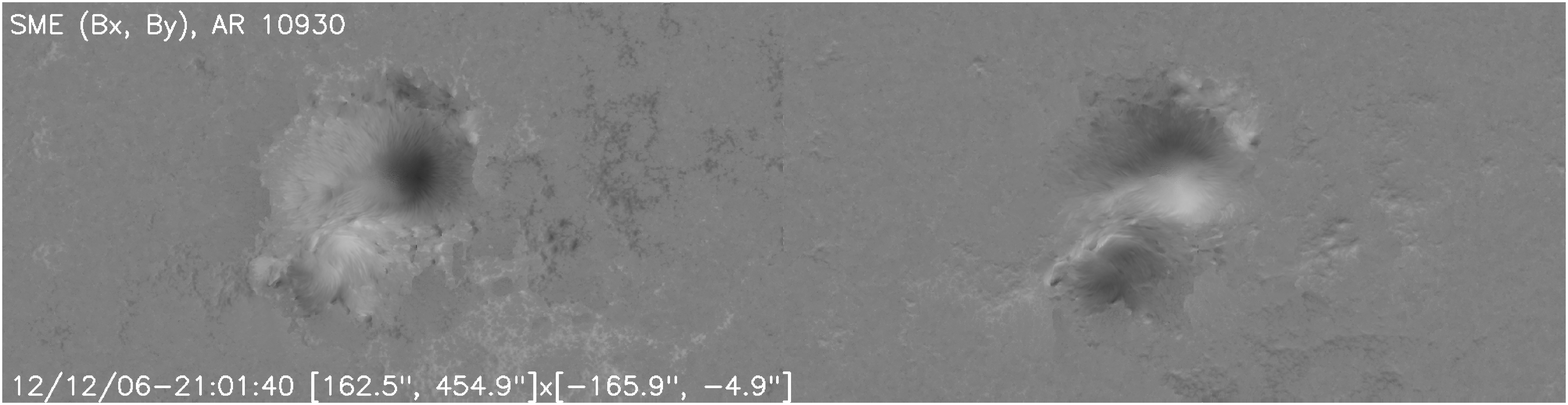}\\
  \caption{Near-Centre SME magnetogram of AR 10930. Lc=19.2}\label{fig:ar10930_1_sme}
\end{figure}

\begin{figure}
  \includegraphics[width=1.0\linewidth]{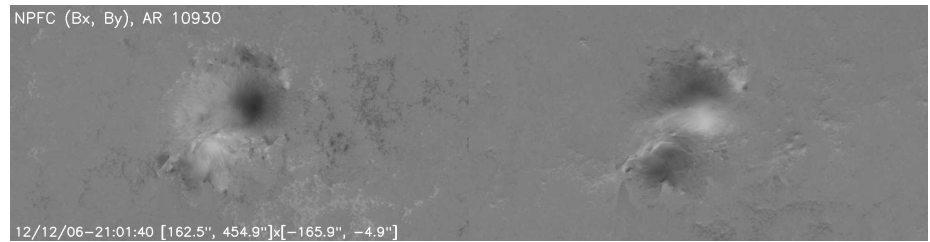}\\
  \caption{Near-Centre NPFC magnetogram of AR 10930. Lc=19.2}\label{fig:ar10930_1_npfc}
\end{figure}

\begin{figure}
  \includegraphics[width=1.0\linewidth]{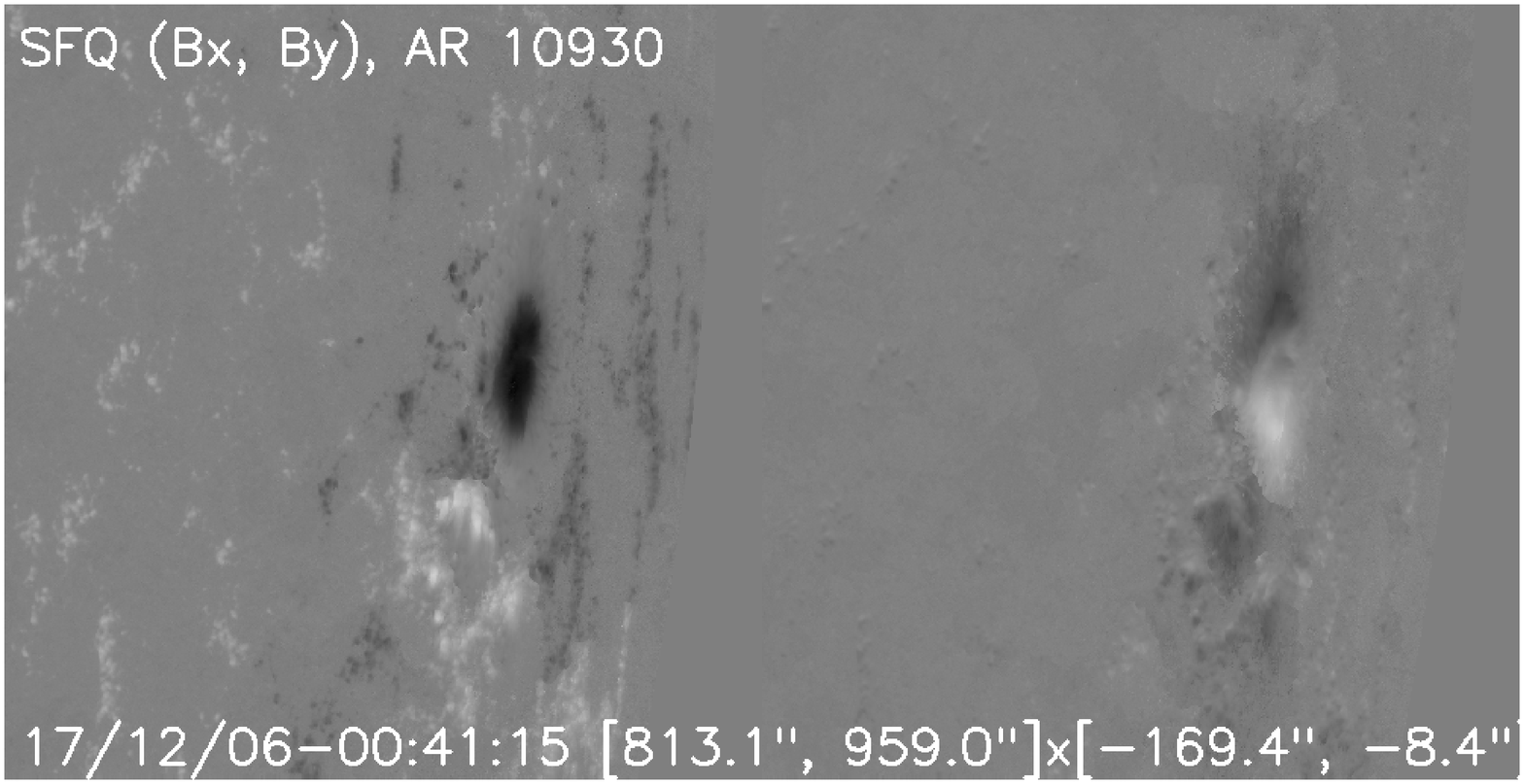}\\
  \caption{Near-Limb SFQ magnetogram of AR 10930. Lc=73.1}\label{fig:ar10930_2}
\end{figure}

\begin{figure}
  \includegraphics[width=1.0\linewidth]{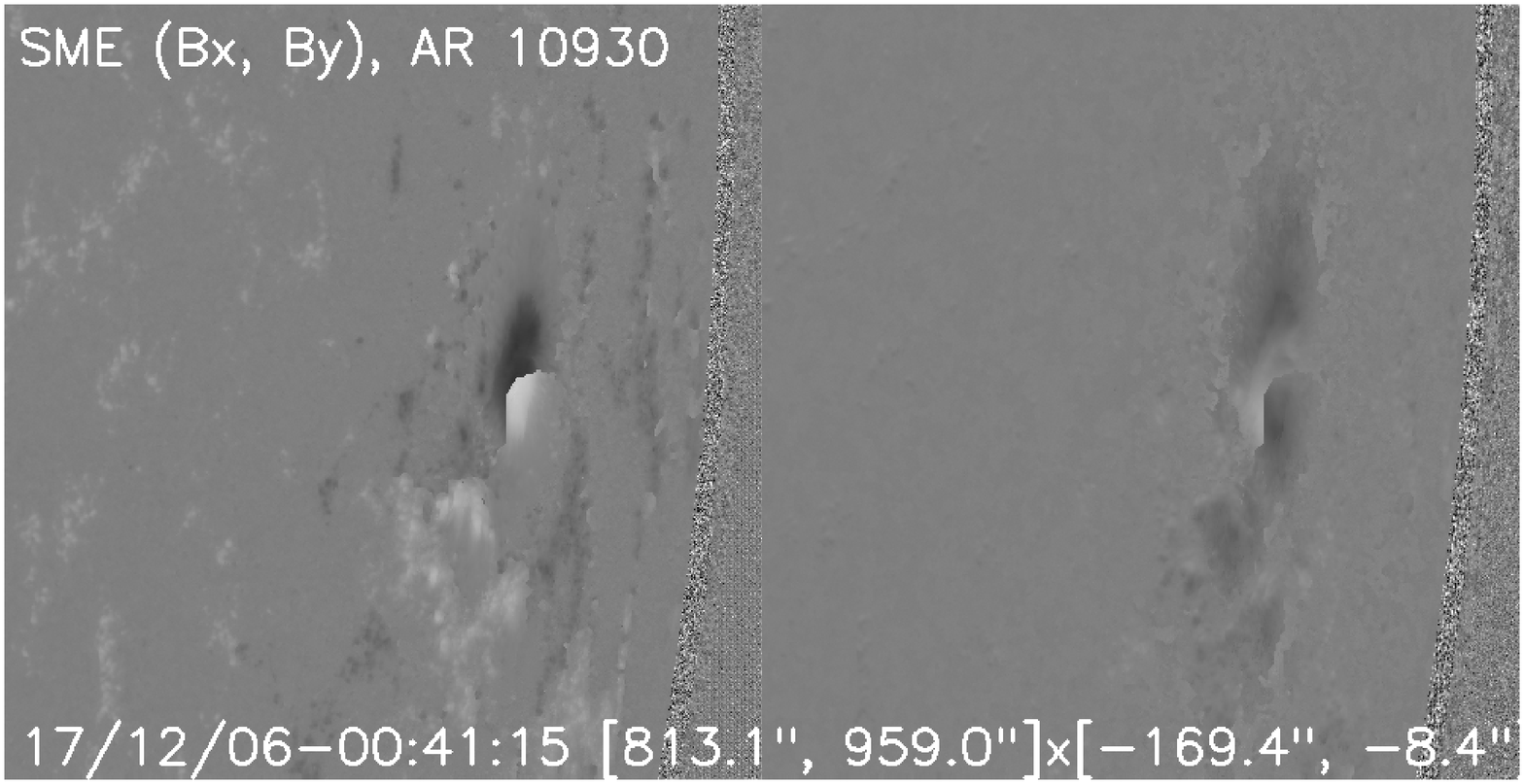}\\
  \caption{Near-Limb SME magnetogram of AR 10930. Lc=73.1}\label{fig:ar10930_2_sme}
\end{figure}

\begin{figure}
  \includegraphics[width=1.0\linewidth]{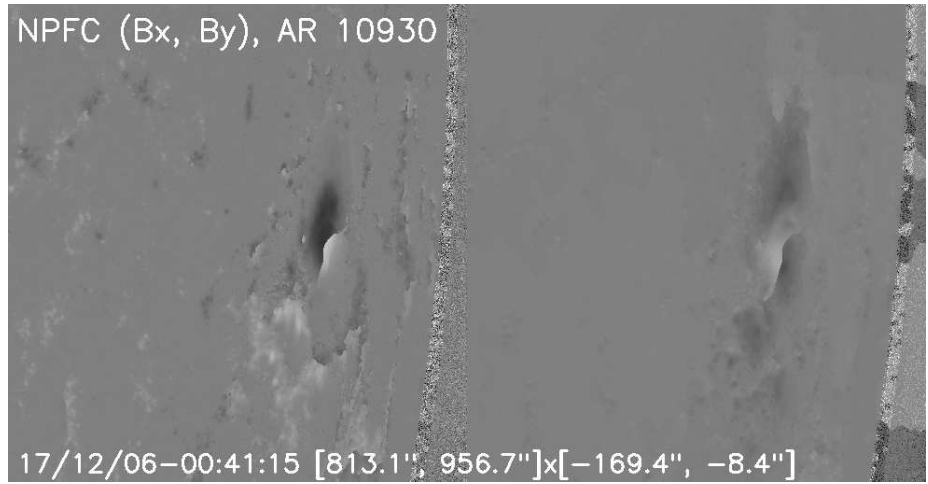}\\
  \caption{Near-Limb NPFC magnetogram of AR 10930. Lc=73.1}\label{fig:ar10930_2_npfc}
\end{figure}

Let us comment on these results, using two cases of the magnetic field location - near the centre (Fig. \ref{fig:ar10930_1}-\ref{fig:ar10930_1_npfc}) and near the limb (Fig. \ref{fig:ar10930_2}-\ref{fig:ar10930_2_npfc}).

Fig.  \ref{fig:ar10930_1}-\ref{fig:ar10930_1_npfc} illustrate
similar quality of   disambiguation by the three methods. When the
magnetic region is at its nearest to the limb (Fig.
\ref{fig:ar10930_2}-\ref{fig:ar10930_2_npfc}), only SFQ code copes
with the task successfully. Throughout this series of
magnetograms, only SFQ code demonstrates satisfactory quality. SME
provides good quality for all magnetograms except for the last
one. NPFC maintains satisfactory level of quality from 12 December
2006, 20:30:05 (Lc=50.1348770), to 15 December 2006, 05:45:05
(Lc=50.1348770). Unlike other codes, SFQ maintains stable level of
quality regardless of distance to the limb. This is consistent
with the quantitative analysis presented in section
\ref{sect:answer}.
\subsection{disambiguation of full-disc SDO/HMI data}
\begin{figure}
  \includegraphics[width=1.0\linewidth]{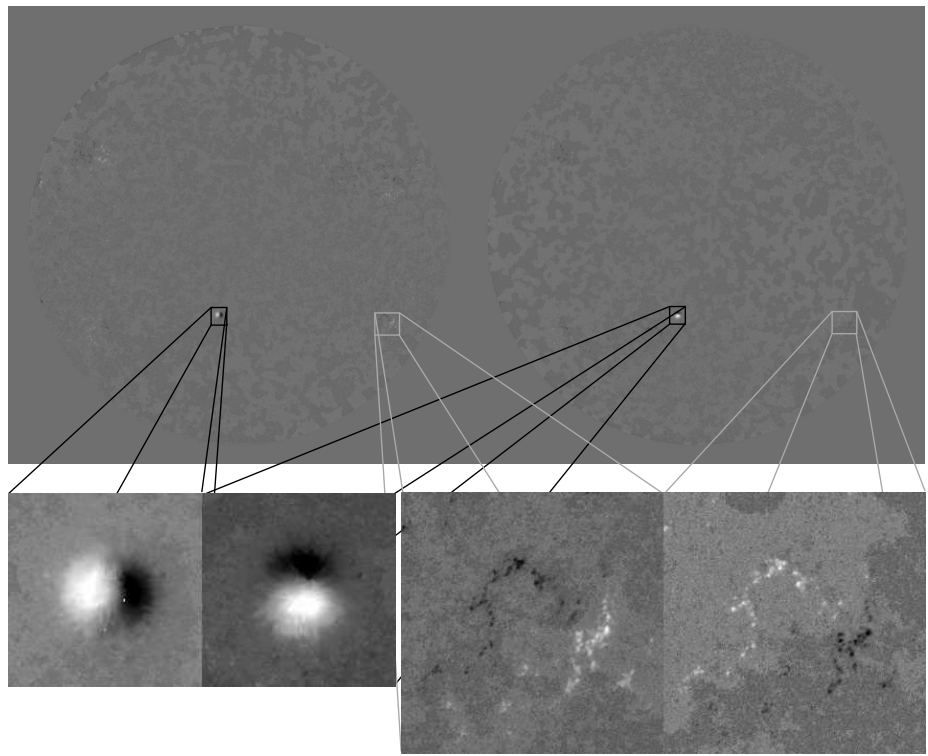}\\
  \caption{Full-disc SDO/HMI on 2 July 2010.}\label{fig:hmi}
\end{figure}

In this subsection, we demonstrate  disambiguation of SDO/HMI data
obtained 2 July 2010 for  the entire disc in original spatial
resolution. Initial data were taken from
\url{http://sun.stanford.edu/~todd /HMICAL/VectorB/}. We divided
data into 10x10 rectangular fragments and sequentially applied the
SFQ code to them. Fig. \ref{fig:hmi} presents the result of
fragment assembly (file containing this image in full resolution
and FITS files of the field components are available on
\url{http://bdm.iszf.irk.ru/SFQ_Disambig/20100702_010000_hmi.zip}).
It takes about one hour to disambiguate  with Core 2 Quad (2.6
GHz) workstation. The result that we have obtained demonstrates
good quality for all large-scale and small-scale magnetic
structures with magnitudes higher than the noise level throughout
the disc. Notice that the corrected magnetogram has a patchwork
structure in the field of very weak fields. This is due to
extremely low signal-to-noise ratio in the regions with very weak
fields

\section{Conclusion}
This paper presents a new code for the azimuth disambiguation of
vector magnetograms. Among  all well-known algorithms for the
azimuth disambiguation, SFQ is the fastest (more than 4 times
faster as compared to the NPFC algorithm). Besides, our algorithm
provides invariably high quality in all parts of the solar disc.
This statement has been proved by testing on well-known analytical
models and using real data from HINODE SOT/SP and SDO/HMI
instruments. In the case of real magnetograms, the method is
efficient near the limb where other algorithms (ME and NPFC) do
not give reliable results. Due to the high speed and quality of
disambiguation, the SFQ method can be applied to process SDO/HMI
vector magnetograms in full resolution (4096x4096) and in almost
real-time mode.

All magnetograms presented in the paper are available on
\url{http://bdm.iszf.irk.ru/SFQ_Disambig}.

\begin{acks}
G. Barnes and NASA/LWS contract NNH05CC75C, K. D. Leka (PI) model
data are used here.

This work was supported by the Ministry of Education and Science
of Russian Federation (GS 8407 and GK 14.518.11.7047) and  by the
RFBR (12-02-31746 mol\_a)
\end{acks}

\begin{table}\caption{$M_{area}$}\label{tab:marea}
    \begin{tabular}{|c|c|c|c|c|}
    \hline
        \textbf{Magnetograms} & \textbf{Lc }& \textbf{SFQ} & \textbf{SME} & \textbf{NPFC}\\ \hline
        20061212\_203005 & 19.2 & 0.80 & 0.93 & 0.74\\ \hline
        20061213\_043005 & 23.5 & 0.80 & 0.93 & 0.74\\ \hline
        20061213\_075005 & 25.2 & 0.80 & 0.93 & 0.74\\ \hline
        20061213\_125104 & 28.0 & 0.81 & 0.91 & 0.74\\ \hline
        20061213\_162104 & 29.8 & 0.80 & 0.91 & 0.75\\ \hline
        20061214\_002005 & 34.1 & 0.81 & 0.89 & 0.74\\ \hline
        20061214\_050005 & 36.7 & 0.81 & 0.88 & 0.75\\ \hline
        20061214\_112602 & 40.3 & 0.81 & 0.87 & 0.74\\ \hline
        20061214\_140103 & 41.6 & 0.80 & 0.86 & 0.74\\ \hline
        20061214\_220005 & 45.9 & 0.81 & 0.84 & 0.74\\ \hline
        20061215\_054505 & 50.1 & 0.80 & 0.82 & 0.74\\ \hline
        20061215\_130205 & 54.2 & 0.80 & 0.81 & 0.73\\ \hline
        20061215\_204604 & 58.3 & 0.81 & 0.79 & 0.72\\ \hline
        20061216\_012106 & 60.8 & 0.80 & 0.78 & 0.72\\ \hline
        20061216\_075005 & 64.4 & 0.80 & 0.78 & 0.70\\ \hline
        20061216\_123104 & 67.0 & 0.80 & 0.78 & 0.69\\ \hline
        20061217\_002528 & 73.1 & 0.80 & 0.76 & 0.56\\ \hline
        20061218\_032421 & 87.9 & 0.76 & - & -\\ \hline
    \end{tabular}
\end{table}
\begin{table}\caption{$M_{flux}$}\label{tab:mflux}
    \begin{tabular}{|c|c|c|c|c|}
    \hline
        \textbf{Magnetograms} & \textbf{Lc }& \textbf{SFQ} & \textbf{SME} & \textbf{NPFC}\\ \hline
     20061212\_203005 & 19.2 & 0.95 & 0.99 & 0.95\\ \hline
    20061213\_043005 & 23.5 & 0.95 & 0.99 & 0.95\\ \hline
    20061213\_075005 & 25.2 & 0.95 & 0.99 & 0.95\\ \hline
    20061213\_125104 & 28.0 & 0.95 & 0.99 & 0.96\\ \hline
    20061213\_162104 & 29.8 & 0.95 & 0.99 & 0.96\\ \hline
    20061214\_002005 & 34.1 & 0.96 & 0.99 & 0.96\\ \hline
    20061214\_050005 & 36.7 & 0.96 & 0.98 & 0.96\\ \hline
    20061214\_112602 & 40.3 & 0.96 & 0.98 & 0.96\\ \hline
    20061214\_140103 & 41.6 & 0.96 & 0.98 & 0.96\\ \hline
    20061214\_220005 & 45.9 & 0.96 & 0.98 & 0.96\\ \hline
    20061215\_054505 & 50.1 & 0.96 & 0.97 & 0.95\\ \hline
    20061215\_130205 & 54.2 & 0.97 & 0.97 & 0.95\\ \hline
    20061215\_204604 & 58.3 & 0.96 & 0.96 & 0.94\\ \hline
    20061216\_012106 & 60.8 & 0.97 & 0.96 & 0.93\\ \hline
    20061216\_075005 & 64.4 & 0.96 & 0.96 & 0.91\\ \hline
    20061216\_123104 & 67.0 & 0.96 & 0.96 & 0.90\\ \hline
    20061217\_002528 & 73.1 & 0.96 & 0.96 & 0.66\\ \hline
    20061218\_032421 & 87.9 & 0.95 & - & -\\ \hline
    \end{tabular}
\end{table}
\begin{table}\caption{$M_{B\bot>500G}$}\label{tab:mbt500}
    \begin{tabular}{|c|c|c|c|c|}
    \hline
        \textbf{Magnetograms} & \textbf{Lc }& \textbf{SFQ} & \textbf{SME} & \textbf{NPFC}\\ \hline
        20061212\_203005 & 19.2 & 0.99 & 1.00 & 0.99\\ \hline
        20061213\_043005 & 23.5 & 0.98 & 1.00 & 0.99\\ \hline
        20061213\_075005 & 25.2 & 0.98 & 1.00 & 0.99\\ \hline
        20061213\_125104 & 28.0 & 0.97 & 1.00 & 0.99\\ \hline
        20061213\_162104 & 29.8 & 0.97 & 1.00 & 0.99\\ \hline
        20061214\_002005 & 34.1 & 0.97 & 1.00 & 0.99\\ \hline
        20061214\_050005 & 36.7 & 0.96 & 1.00 & 0.99\\ \hline
        20061214\_112602 & 40.3 & 0.96 & 1.00 & 0.99\\ \hline
        20061214\_140103 & 41.6 & 0.96 & 1.00 & 0.99\\ \hline
        20061214\_220005 & 45.9 & 0.96 & 1.00 & 0.99\\ \hline
        20061215\_054505 & 50.1 & 0.96 & 1.00 & 0.99\\ \hline
        20061215\_130205 & 54.2 & 0.97 & 1.00 & 0.99\\ \hline
        20061215\_204604 & 58.3 & 0.97 & 1.00 & 0.98\\ \hline
        20061216\_012106 & 60.8 & 0.97 & 1.00 & 0.98\\ \hline
        20061216\_075005 & 64.4 & 0.96 & 0.99 & 0.96\\ \hline
        20061216\_123104 & 67.0 & 0.96 & 0.99 & 0.95\\ \hline
        20061217\_002528 & 73.1 & 0.97 & 0.99 & 0.61\\ \hline
        20061218\_032421 & 87.9 & 0.95 & - & -\\
    \hline
    \end{tabular}
\end{table}
\begin{table}\caption{$M_{B\bot>100G}$}\label{tab:mbt100}
    \begin{tabular}{|c|c|c|c|c|}
    \hline
        \textbf{Magnetograms} & \textbf{Lc }& \textbf{SFQ} & \textbf{SME} & \textbf{NPFC}\\ \hline
        20061212\_203005 & 19.2 & 0.96 & 0.99 & 0.95\\ \hline
    20061213\_043005 & 23.5 & 0.96 & 0.99 & 0.95\\ \hline
    20061213\_075005 & 25.2 & 0.95 & 0.99 & 0.95\\ \hline
    20061213\_125104 & 28.0 & 0.96 & 0.99 & 0.95\\ \hline
    20061213\_162104 & 29.8 & 0.96 & 0.99 & 0.96\\ \hline
    20061214\_002005 & 34.1 & 0.96 & 0.99 & 0.96\\ \hline
    20061214\_050005 & 36.7 & 0.96 & 0.99 & 0.96\\ \hline
    20061214\_112602 & 40.3 & 0.95 & 0.99 & 0.96\\ \hline
    20061214\_140103 & 41.6 & 0.95 & 0.99 & 0.96\\ \hline
    20061214\_220005 & 45.9 & 0.96 & 0.99 & 0.97\\ \hline
    20061215\_054505 & 50.1 & 0.96 & 0.99 & 0.97\\ \hline
    20061215\_130205 & 54.2 & 0.96 & 0.99 & 0.96\\ \hline
    20061215\_204604 & 58.3 & 0.96 & 0.98 & 0.96\\ \hline
    20061216\_012106 & 60.8 & 0.96 & 0.98 & 0.95\\ \hline
    20061216\_075005 & 64.4 & 0.95 & 0.97 & 0.93\\ \hline
    20061216\_123104 & 67.0 & 0.95 & 0.97 & 0.93\\ \hline
    20061217\_002528 & 73.1 & 0.95 & 0.96 & 0.62\\ \hline
    20061218\_032421 & 87.9 & 0.93 & - & -\\
    \hline
    \end{tabular}
\end{table}
\begin{table}\caption{$M_{B\Delta}(G)$}\label{tab:mbdelta}
    \begin{tabular}{|c|c|c|c|c|}
    \hline
        \textbf{Magnetograms} & \textbf{Lc }& \textbf{SFQ} & \textbf{SME} & \textbf{NPFC}\\ \hline
        20061212\_203005 & 19.2 & 28.3 & 8.1 & 38.2\\ \hline
        20061213\_043005 & 23.5 & 27.6 & 8.2 & 37.4\\ \hline
        20061213\_075005 & 25.2 & 28.4 & 6.7 & 38.0\\ \hline
        20061213\_125104 & 28.0 & 25.3 & 9.3 & 38.0\\ \hline
        20061213\_162104 & 29.8 & 27.1 & 8.2 & 36.2\\ \hline
        20061214\_002005 & 34.1 & 25.8 & 10.8 & 36.2\\ \hline
        20061214\_050005 & 36.7 & 24.7 & 11.9 & 35.5\\ \hline
        20061214\_112602 & 40.3 & 25.4 & 13.0 & 36.3\\ \hline
        20061214\_140103 & 41.6 & 29.1 & 14.6 & 36.8\\ \hline
        20061214\_220005 & 45.9 & 25.8 & 17.0 & 36.0\\ \hline
        20061215\_054505 & 50.1 & 27.1 & 19.9 & 36.6\\ \hline
        20061215\_130205 & 54.2 & 25.6 & 21.5 & 36.0\\ \hline
        20061215\_204604 & 58.3 & 27.3 & 25.0 & 45.6\\ \hline
        20061216\_012106 & 60.8 & 26.7 & 27.7 & 46.2\\ \hline
        20061216\_075005 & 64.4 & 28.6 & 26.5 & 62.1\\ \hline
        20061216\_123104 & 67.0 & 29.5 & 26.9 & 70.1\\ \hline
        20061217\_002528 & 73.1 & 28.8 & 29.8 & 257.8\\ \hline
        20061218\_032421 & 87.9 & 32.6 & - & -\\ \hline
    \end{tabular}
\end{table}
\begin{table}\caption{$M_{Jz}$}\label{tab:mjz}
    \begin{tabular}{|c|c|c|c|c|}
    \hline
        \textbf{Magnetograms} & \textbf{Lc }& \textbf{SFQ} & \textbf{SME} & \textbf{NPFC}\\ \hline
        20061212\_203005 & 19.2 & 0.82 & 0.95 & 0.76\\ \hline
        20061213\_043005 & 23.5 & 0.83 & 0.95 & 0.76\\ \hline
        20061213\_075005 & 25.2 & 0.83 & 0.95 & 0.76\\ \hline
        20061213\_125104 & 28.0 & 0.84 & 0.94 & 0.76\\ \hline
        20061213\_162104 & 29.8 & 0.83 & 0.95 & 0.77\\ \hline
        20061214\_002005 & 34.1 & 0.83 & 0.93 & 0.78\\ \hline
        20061214\_050005 & 36.7 & 0.84 & 0.92 & 0.78\\ \hline
        20061214\_112602 & 40.3 & 0.84 & 0.92 & 0.78\\ \hline
        20061214\_140103 & 41.6 & 0.82 & 0.91 & 0.78\\ \hline
        20061214\_220005 & 45.9 & 0.84 & 0.90 & 0.79\\ \hline
        20061215\_054505 & 50.1 & 0.84 & 0.89 & 0.79\\ \hline
        20061215\_130205 & 54.2 & 0.85 & 0.89 & 0.80\\ \hline
        20061215\_204604 & 58.3 & 0.85 & 0.88 & 0.77\\ \hline
        20061216\_012106 & 60.8 & 0.86 & 0.87 & 0.77\\ \hline
        20061216\_075005 & 64.4 & 0.86 & 0.87 & 0.75\\ \hline
        20061216\_123104 & 67.0 & 0.86 & 0.87 & 0.74\\ \hline
        20061217\_002528 & 73.1 & 0.87 & 0.86 & 0.55\\ \hline
        20061218\_032421 & 87.9 & 0.90 & - & -\\ \hline
    \end{tabular}
\end{table}


\end{article}

\end{document}